\documentclass[a4paper,12pt]{article}
\usepackage[english]{babel}
\usepackage{graphicx,color}
\usepackage{amsmath,amssymb,amsfonts,bm,euscript,amscd,ae,amsthm,exscale,amsopn,amstext}
\usepackage{ifthen}
\usepackage[T1]{fontenc}
\usepackage[cp1251]{inputenc}
\usepackage[active]{srcltx}

\title{Novel Solution of Wheeler--DeWitt Theory}

\author{\textbf{Lukasz Andrzej Glinka}\footnote{E-mail address: laglinka@gmail.com, lukaszglinka@wp.eu}}
\date{\empty}

\begin{document}
\maketitle

\begin{abstract}
Taking into account the global one-dimensionality conjecture recently proposed by the author, the Cauchy-like analytical wave functional of the Wheeler--DeWitt theory is derived. The crucial point of the integration strategy is canceling of the singular behavior of the effective potential, which is performed through the suitable change of variables introducing the invariant global dimension. In addition, the conjecture is extended onto the wave functionals dependent on both Matter fields as well as the invariant global dimension. Through application of the reduction within the quantum gravity model, the appropriate Dirac equation is obtained and than solved. The case of superposition is also briefly discussed.\\

\end{abstract}

\newpage
\section{Introduction}

The Wheeler--DeWitt theory, also known as quantum geometrodynamics or quantum General Relativity, is the foundational model of quantum gravity considered in modern theoretical physics, Cf. Ref. \cite{qg}. This model straightforwardly arises from the canonical General Relativity, formulated on the basis of the Arnowitt--Deser--Misner decomposition, well known as the $3+1$ splitting, of a four-dimensional space-time metric, applied to the Einstein--Hilbert action supplemented by the York--Gibbons--Hawking boundary term. This procedure produces the Hamiltonian action, as well as the primary and secondary constraints satisfying the first-class algebra whose are canonically quantized according to the Dirac method. The quantized Hamiltonian constraint is the quantum evolution known as the Wheeler-DeWitt equation, which is the second order functional differential equation on the abstract configuration space known as the Wheeler superspace and containing all three-dimensional embedded geometries, whose solutions known as wave functionals in general depend on an induced three-metric and Matter fields.

The heart of the matter, however, is the question of integrability of the Wheeler--DeWitt equation, and, for this reason, the possible new physical meaning of the quantum geometrodynamics could arise along with the integration strategy. Since the 1970s, S. W. Hawking and his coauthors \cite{haw} have proposed to solve the Wheeler-DeWitt equation through making use of the formal analogy with the Schr\"odinger equation of usual quantum mechanics, and applied the Feynman path integral method which, however, generates manifestly non--analytical wave functionals, that is the solutions which do not form the Cauchy surface necessary to the rational analysis of any differential equation. The approach, sometimes called the Hartle--Hawking wave function, is correct from the point of view of quantum field theory, but actually instead of concrete calculations of the path integrals and development of the method beyond the simplest cosmological solutions of the Einstein field equations, the only qualitative ideas which link the Feynman integration with quantum cosmology have been proposed. Since this approach is far from a mathematical consistency, still the question which is neglected in the literature are other possible solutions to the Wheeler-DeWitt equation, including both non-analytical and analytical ones, which could be beyond the solutions defined through the method of path integration. It is worth stressing that, in fact, both any one analytical solution, that is the Cauchy-like wave functional, to quantum geometrodynamics and its possibly interesting physical meaning are still unknown. This point is very unsatisfactory, and, consequently, makes the quantum geometrodynamics a theory produced in the way of a false analogy with the formalism of quantum mechanics.

Nevertheless, the discussion of the qualitative matter of quantum geometrodynamics is not the subject of this paper, whereas we will present here the way to receive the analytical solutions to the Wheeler--DeWitt equation throughout a systematic construction. For this reason, we apply the global one-dimensional conjecture, recently discussed by the author's writings \cite{gli}, which is immediately rooted in the generic quantum cosmology \cite{generic} dedicated to the Einstein-Friedmann Universe. The main result of this conjecture are the wave functional which are dependent on the determinant of the three-dimensional metric, which is named the global dimension, and the resulting theory is the Schr\"odinger quantum mechanics in the one dimension. The crucial point of the integration strategy is application of the suitable change of variables which removes the singular behavior of the effective potential. In result, we receive the concept of invariant global dimension, where the word invariance is related to the invariant integral measure on a spacelike hypersurface, and the presence of Matter fields is included. Finally, we show the way to transform the theory into the suitable Dirac equation, which defined the new strategy for quantum geometrodynamics and is solved to receive the analytical wave functionals.

The paper is organized in the following way. In the Section \ref{sec:1}, the basic facts about quantum geometrodynamics are collected. The Section \ref{sec:2} briefly discusses the global one-dimensional conjecture, including the concept of the invariant global dimension. The suitable Dirac equation is obtained in the Section \ref{sec:3}, and the new type of analytical wave functionals is constructed in the Section \ref{sec:4}. In Section \ref{sec:5}, certain consequences are presented, whereas in the Section \ref{sec:6} all results are summarized.

\section{\mbox{Canonical Quantum Gravity}}\label{sec:1}
Let us recall the basic facts, for details Cf. Ref. \cite{4dim}. General Relativity, governed by the Einstein field equations\footnote{In this paper we use the units in units $8\pi G/3=1$, $c=1$, $\hslash=1$.}
\begin{equation}\label{feq}
R_{\mu\nu}-\dfrac{1}{2}g_{\mu\nu}{^{(4)}\!}R+\Lambda g_{\mu\nu}=3T_{\mu\nu},
\end{equation}
where $\Lambda$ is a cosmological constant and $T_{\mu\nu}$ is a stress-energy tensor of Matter fields, models space-time as a four-dimensional pseudo--Riemannian manifold $(M,g)$ equipped with a metric $g_{\mu\nu}$, the Riemann--Christoffel curvature tensor $R^\lambda_{\mu\alpha\nu}$, the Ricci second fundamental form $R_{\mu\nu}=R^\lambda_{\mu\lambda\nu}$, and the Ricci scalar curvature ${^{(4)}\!}R=g^{\kappa\lambda}R_{\kappa\lambda}$. If $M$ is closed and has an induced spacelike boundary $(\partial M,h)$ with an induced metric $h_{ij}$, the second fundamental form $K_{ij}$, and an extrinsic curvature $K=h^{ij}K_{ij}$, then the field equations (\ref{feq}) are the equations of motion following from the variational principle applied to the Einstein--Hilbert action with the York--Gibbons--Hawking term \cite{ygh}
\begin{equation}\label{eh0}
S[g]\!=\!\int_{M}d^4x\sqrt{-g}\left\{-\dfrac{1}{6}{^{(4)}\!}R+\dfrac{\Lambda}{3}\right\}+S_\phi[g]-\dfrac{1}{3}\int_{\partial M}d^3x\sqrt{h}K\quad,
\end{equation}
and the stress--energy tensor generated by the variational principle is
\begin{equation}
  T_{\mu\nu}=-\dfrac{2}{\sqrt{-g}}\dfrac{\delta S_\phi[g]}{\delta g^{\mu\nu}}\quad,\quad S_\phi[g]\equiv\int_M d^4x\sqrt{-g}L_\phi,
\end{equation}
where $L_\phi$ is Matter fields Lagrangian. The appropriate embedding theorems allow to make use of the $3+1$ splitting \cite{adm}
\begin{eqnarray}\label{dec}
g_{\mu\nu}=\left[\begin{array}{cc}-N^2+N^iN_i&N_j\\N_i&h_{ij}\end{array}\right]\quad,\quad h_{ik}h^{kj}=\delta_i^j\quad,\quad N^i=h^{ij}N_j,
\end{eqnarray}
for which the action (\ref{eh0}) takes the Hamiltonian form $S[g]=\int dt L$ with
\begin{eqnarray}\label{gd}
L=\int_{\partial M} d^3x\left\{\pi_\phi\dot{\phi}+\pi\dot{N}+\pi^i\dot{N_i}+\pi^{ij}\dot{h}_{ij}-NH-N_iH^i\right\},
\end{eqnarray}
 where $\pi$'s are canonical conjugate momenta, and $H$, $H^i$ are \cite{gd}
\begin{eqnarray}
\!\!\!\!\!\!\!\!\!\!\!\!\!\!\!&\pi_\phi=\frac{\partial L_\phi}{\partial \dot{\phi}}\quad,\quad \pi=\frac{\partial L}{\partial \dot{N}}\quad,\quad\pi^i=\frac{\partial L}{\partial \dot{N_i}}\quad,\quad\pi^{ij}=\sqrt{h}\left(K^{ij}-Kh^{ij}\right),&\label{con1}\\
\!\!\!\!\!\!\!\!\!\!\!\!\!\!\!&H^i=2\pi^{ij}_{~;j}\quad,\quad H=\sqrt{h}\left\{{^{(3)}\!R}[h]+K^2-K_{ij}K^{ij}-2\Lambda-6\varrho[\phi]\right\},&\label{con2}
\end{eqnarray}
with ${^{(3)}\!R}\equiv h^{ij}R_{ij}$, $\varrho[\phi]=n^{\mu}n^{\nu}T_{\mu\nu}$, $n^\mu=(1/N)\left[1,-N^i\right]$, and holds
\begin{equation}
  \dot{h}_{ij}=2NK_{ij}+N_{i|j}+N_{j|i}.\label{con0}
\end{equation}
where $N_{i|j}$ is an intrinsic covariant derivative of $N_i$. $H^i$ are generators of the spatial dif\/feomorphisms $\widetilde{x}^i=x^i+\xi^i$ \cite{dew}
\begin{eqnarray}
i\left[h_{ij},\int_{\partial M}H_{a}\xi^a d^3x\right]&=&-h_{ij,k}\xi^k-h_{kj}\xi^{k}_{~,i}-h_{ik}\xi^{k}_{~,j}~~,\\
i\left[\pi^{ij},\int_{\partial M}H_{a}\xi^a d^3x\right]&=&-\left(\pi^{ij}\xi^k\right)_{,k}+\pi^{kj}\xi^{i}_{~,k}+\pi^{ik}\xi^{j}_{~,k}~~,
\end{eqnarray}
where $H_i=h_{ij}H^j$, and the first-class algebra holds
\begin{eqnarray}
\!\!\!\!\!\!\!\!\!\!&&i\left[H_i(x),H_j(y)\right]=\int_{\partial M}H_{a}c^a_{ij}d^3z\quad,\quad i\left[H(x),H_i(y)\right]=H\delta^{(3)}_{,i}(x,y),\label{com2}\\
\!\!\!\!\!\!\!\!\!\!&&i\left[\int_{\partial M}H\xi_1d^3x,\int_{\partial M}H\xi_2d^3x\right]=\int_{\partial M}H^a\left(\xi_{1,a}\xi_2-\xi_1\xi_{2,a}\right)d^3x,\label{com3}
\end{eqnarray}
where $c^a_{ij}=\delta^a_i\delta^b_j\delta^{(3)}_{,b}(x,z)\delta^{(3)}(y,z)-(i\leftrightarrow j,x\leftrightarrow y)$ are the structure constants of the dif\/feomorphism group, and all other Lie's brackets vanish. Time-preservation \cite{dir} of the primary constraints, that is $\pi\approx0$ and $\pi^i\approx0$, leads to the secondary constraints - scalar (Hamiltonian) and vector respectively
\begin{eqnarray}
H\approx0\quad, \quad H^i\approx0\quad,
\end{eqnarray}
where the scalar constraint yields dynamics, while the vector one merely reflects dif\/feoinvariance. Making use of the canonical momentum $\pi^{ij}$, one obtains the Einstein--Hamilton--Jacobi equation
\begin{equation}\label{con}
H=G_{ijkl}\pi^{ij}\pi^{kl}-\sqrt{h}\left(^{(3)}R[h]-2\Lambda-6\varrho[\phi]\right)\approx0\quad,
\end{equation}
where $G_{ijkl}\equiv(2\sqrt{h})^{-1}\left(h_{ik}h_{jl}+h_{il}h_{jk}-h_{ij}h_{kl}\right)$ is the DeWitt metric on the Wheeler superspace \cite{sup}. The Dirac quantization method \cite{dir}
\begin{eqnarray}\label{dpq}
i\left[\pi^{ij}(x),h_{kl}(y)\right]&=&\dfrac{1}{2}\left(\delta_{k}^{i}\delta_{l}^{j}+\delta_{l}^{i}\delta_{k}^{j}\right)\delta^{(3)}(x,y)\quad,\\
i\left[\pi^i(x),N_j(y)\right]&=&\delta^i_j\delta^{(3)}(x,y)\quad,\quad i\left[\pi(x),N(y)\right]=\delta^{(3)}(x,y)\quad,
\end{eqnarray}
applied to the constraint (\ref{con}), yields the Wheeler--DeWitt equation \cite{whe, dew}
\begin{equation}\label{wdw}
\left\{G_{ijkl}\dfrac{\delta^2}{\delta h_{ij}\delta h_{kl}}+\sqrt{h}\left({^{(3)}\!R}[h]-2\Lambda-6\varrho[\phi]\right)\right\}\Psi[h_{ij},\phi]=0\quad,
\end{equation}
whereas other first class constraints merely reflect dif\/feoinvariance
\begin{equation}\label{diff}
  \pi\Psi[h_{ij},\phi]=0\quad, \quad \pi^i\Psi[h_{ij},\phi]=0\quad, \quad H^i\Psi[h_{ij},\phi]=0\quad,
\end{equation}
and are not important in this model, called quantum geometrodynamics.
\section{Global one-dimensional conjecture}\label{sec:2}

The global one--dimensionality conjecture \cite{gli}, establishes the strategy within quantum geometrodynamics which allows to receive analytical solutions. Making use of the Jacobi rule for differentiation of a determinant of a metric $g_{\mu\nu}$ one obtains
\begin{equation}
  \delta g=gg^{\mu\nu}\delta g_{\mu\nu} \longrightarrow N^2\delta h=N^2hh^{ij}\delta h_{ij},
\end{equation}
where $h=\det h_{ij}=\frac{1}{3}\epsilon^{ijk}\epsilon^{abc}h_{ia}h_{jb}h_{kc}$ is the diffeoinvariant variable which is third order in $h_{ij}$, and $\epsilon^{ijk}$ is the Levi-Civita density. Consequently, one has the differentiation rule
\begin{equation}\label{hhij}
  \dfrac{\delta}{\delta h_{ij}}\Psi[h_{ij},\phi]=hh^{ij}\dfrac{\delta}{\delta h}\Psi[h,\phi]\quad,
\end{equation}
which applied to the quantum geometrodynamics (\ref{wdw}), with making the double contraction of the supermetric with an embedding metric, leads to
\begin{equation}
G_{ijkl}\dfrac{\delta^2}{\delta h_{ij}\delta h_{kl}}\Psi[h_{ij},\phi]=-\dfrac{3}{2}h^{3/2}\dfrac{\delta^2}{\delta h^2}\Psi[h,\phi],
\end{equation}
and finally the Wheeler--DeWitt equation (\ref{wdw}) becomes the usual differential equation
\begin{equation}\label{kgf}
\left(\dfrac{\delta^2}{\delta{h^2}}+V_{eff}[h,\phi]\right)\Psi[h, \phi]=0,
\end{equation}
where $V_{eff}[h,\phi]$ is the ef\/fective potential
\begin{equation}\label{eff}
  V_{eff}[h,\phi]\equiv \dfrac{2}{3}\dfrac{{^{(3)}\!R}}{h}-\dfrac{4}{3}\dfrac{\Lambda}{h}-\dfrac{4}{h}\varrho[\phi].
\end{equation}
The first term in (\ref{eff}) describes contribution due to an embedding geometry only, the second one is mix of the cosmological constant and an embedding geometry, and the third component is due to Matter fields and an embedding geometry. In result, one has to deal with wave functionals $\Psi[h_{ij},\phi] = \Psi[h,\phi]$, what agrees with the basic diffeoinvariace (\ref{diff}).

The potential (\ref{eff}) has a manifestly singular behavior $\sim1/h$, which however can be canceled through the appropriate change of variables
\begin{eqnarray}\label{kgfrd}
&&h\rightarrow \xi=\xi[h],\\
&&\dfrac{\delta \xi}{\delta h}\neq0,\\
&&\delta\xi=\left(\dfrac{\delta \xi}{\delta h}\right)hh^{ij}\delta h_{ij},
\end{eqnarray}
where we have introduced the new global dimension $\xi[h]$, called here the invariant dimension, which is a functional of the global dimension $h$ and, therefore, also diffeoinvariant. With (\ref{kgfrd}) the equation (\ref{kgf}) becomes
\begin{equation}\label{kgfr}
\left\{\dfrac{\delta^2}{\delta{\xi^2}}+V[\xi,\phi]\right\}\Psi\left[\xi,\phi\right]=0,
\end{equation}
where
\begin{equation}
  V[\xi,\phi]=\left(\dfrac{\delta \xi}{\delta h}\right)^{-2}V_{eff}\left[\xi,\phi\right].
\end{equation}
In fact, $\xi$ is a kind of the gauge, wherein $\xi[h]\equiv h$ is generic. Note that the following choice
\begin{eqnarray}
  \xi&=&\sqrt{\dfrac{8}{3}}\sqrt{\strut{h}},\label{xi1}\\
  \delta\xi&=&\sqrt{\dfrac{2}{3}}\dfrac{\delta h}{\sqrt{h}}=\sqrt{\dfrac{2}{3}}\sqrt{\strut{h}}h^{ij}\delta h_{ij}\label{xi2},
\end{eqnarray}
cancels the singularity $1/h$ in $V_{eff}\left[h,\phi\right]$ (\ref{eff}), and the equation (\ref{kgfr}) becomes
\begin{equation}\label{eqn}
\left\{\dfrac{\delta^2}{\delta{\xi^2}}+{^{(3)}\!R[\xi]}-2\Lambda-6\varrho[\phi]\right\}\Psi\left[\xi,\phi\right]=0,
\end{equation}
with the appropriate normalization condition
\begin{equation}\label{norm}
  \int\left|\Psi\left[\xi,\phi\right]\right|^2\delta\mu(\xi,\phi)=1,
\end{equation}
where $\delta\mu(\xi,\phi)=\delta\xi\delta\phi$ is the invariant product functional measure. Note that both $\delta h$ and $\delta \sqrt{h}$ are the Lebesgue--Stieltjes (Radon) integral measures which can be rewritten as the Riemann measures
\begin{equation}
\delta \sqrt{h} = \dfrac{\partial^4 \sqrt{h}}{\partial x_0\partial x_1\partial x_2\partial x_3}d^4x,\quad h=h(x_0,x_1,x_2,x_3)
\end{equation}
what relates the superspace to the space-time.
\section{The Dirac equation}\label{sec:3}
Eq. (\ref{kgfr}) can be derived as the Euler-Lagrange equation of motion by variational principle $\delta S[\Psi]=0$ applied to the action
\begin{eqnarray}
\!\!\!\!\!\!\!\!\!\!\!\!\!\!\!&&S[\Psi]=-\dfrac{1}{2}\int \delta \xi\delta\phi\Psi[\xi,\phi]\left(\dfrac{\delta^2}{\delta{\xi^2}}+V[\xi,\phi]\right)\Psi[\xi,\phi]=\\
\!\!\!\!\!\!\!\!\!\!\!\!\!\!\!&&=-\dfrac{1}{2}\int\delta\phi\Psi[\xi,\phi]\dfrac{\delta\Psi[\xi,\phi]}{\delta\xi}+\dfrac{1}{2}\int\delta\xi\delta\phi \left\{\left(\dfrac{\delta \Psi[\xi,\phi]}{\delta \xi}\right)^2+V[\xi,\phi]\Psi^2[\xi,\phi]\right\},
\end{eqnarray}
where partial differentiation was used. Choosing the coordinate system so that the boundary term vanishes
\begin{equation}\label{choice}
-\dfrac{1}{2}\int\delta\phi\Psi[\xi,\phi]\dfrac{\delta\Psi[\xi,\phi]}{\delta\xi}=0,
\end{equation}
and using the standard definition
\begin{equation}
S[\Psi]\equiv\int\delta \xi\delta\phi L\left[\Psi[\xi,\phi],\delta \Psi[\xi,\phi]/\delta \xi\right],
\end{equation}
one obtains the Lagrangian of the Euclidean field theory
\begin{equation}
  L\left[\Psi[\xi,\phi],\dfrac{\delta \Psi[\xi,\phi]}{\delta \xi}\right]=\dfrac{1}{2}\left(\dfrac{\delta \Psi[\xi,\phi]}{\delta \xi}\right)^2+\dfrac{V[\xi,\phi]}{2}\Psi^2[\xi,\phi],
\end{equation}
for which the corresponding canonical conjugate momentum is
\begin{equation}\label{1}
  \Pi_\Psi[\xi,\phi]=\dfrac{\partial L}{\partial \left(\delta \Psi[\xi,\phi]/\delta \xi\right)}=\dfrac{\delta \Psi[\xi,\phi]}{\delta \xi},
\end{equation}
and, therefore, the choice (\ref{choice}) actually means orthogonal coordinates
\begin{equation}\label{coord}
  \Psi[\xi,\phi]\Pi_{\Psi}[\xi,\phi]=0,
\end{equation}
for any values of $\xi$ and $\phi$. Applying (\ref{1}) in (\ref{kgfr}), one receives
\begin{equation}\label{2}
  \dfrac{\delta\Pi_\Psi[\xi,\phi]}{\delta \xi}+V[\xi,\phi]\Psi[\xi,\phi]=0,
\end{equation}
and combining with (\ref{1}), the appropriate Dirac equation is obtained
\begin{equation}\label{dira}
 \left(i\gamma\dfrac{\delta}{\delta \xi}-M[\xi,\phi]\right)\Phi[\xi,\phi]=0,
\end{equation}
where we have employed the notation
\begin{equation}
  \Phi[\xi,\phi]=\left[\begin{array}{c}\Pi_\Psi[\xi,\phi]\\ \Psi[\xi,\phi]\end{array}\right]\quad,\quad M[\xi,\phi]=\left[\begin{array}{cc}
1&0\\0&V[\xi,\phi]\end{array}\right],
\end{equation}
and the $\gamma$-matrices algebra consists only one element - the Pauli matrix $\sigma_y$
\begin{equation}
  \gamma=\left[\begin{array}{cc}0&-i\\i&0\end{array}\right]\equiv\sigma_y\quad,\quad\gamma^2=I,
\end{equation}
where $I$ is the identity matrix, that in itself obey the algebra
\begin{equation}
  \left\{\gamma,\gamma\right\}=2\delta_E\quad,\quad\delta_E=\left[\begin{array}{cc}1&0\\0&1\end{array}\right].
\end{equation}
Dimensional reduction of the one component second order theory (\ref{kgfr}) yields the two component first order one (\ref{dira}) determined by the Eucludean Clifford algebra $\mathcal{C}\ell_{1,1}(\mathbb{R})$, Cf. Ref. \cite{euc}, that is the matrix algebra having a complex two-dimensional representation, which decomposes into a direct sum of two isomorphic central simple algebras or a tensor product
\begin{eqnarray}
\!\!\!\!\!\!\!\!\!\!&&\mathcal{C}\ell_{1,1}(\mathbb{R})=\mathcal{C}\ell^+_{1,1}(\mathbb{R})\oplus \mathcal{C}\ell^-_{1,1}(\mathbb{R})=\mathcal{C}\ell_{2,0}(\mathbb{R})\otimes\mathcal{C}\ell_{0,0}(\mathbb{R}),~~~~~~\\
\!\!\!\!\!\!\!\!\!\!&&\mathcal{C}\ell_{1,1}(\mathbb{R})\cong\mathbb{R}(2)\quad,\quad\mathcal{C}\ell^{\pm}_{1,1}(\mathbb{R})=\dfrac{1\pm\gamma}{2}\mathcal{C}\ell_{1,1}(\mathbb{R})\cong\mathbb{R}\quad,\quad\mathcal{C}\ell_{0,0}(\mathbb{R})\cong\mathbb{R}.~~~~~~
\end{eqnarray}
 Restricting to $Pin_{1,1}(\mathbb{R})$ yield a two-dimensional spin representations; $Spin_{1,1}(\mathbb{R})$ splits it onto a sum of two one-dimensional Weyl representations.
\section{Analytic wave functional}\label{sec:4}
The Dirac equation (\ref{dira}) can be rewritten in the Schr\"odinger form
\begin{equation}\label{evol}
  i\dfrac{\delta\Phi[\xi,\phi]}{\delta \xi}=H[\xi,\phi]\Phi[\xi,\phi]\quad,\quad H[\xi,\phi]=i\left[\begin{array}{cc}0&-V[\xi,\phi]\\1&0\end{array}\right],
\end{equation}
whose most general solution can be written as
\begin{equation}
  \Phi[\xi,\phi]=U[\xi,\phi]\Phi[\xi^I,\phi],
\end{equation}
where $\Phi[\xi^I,\phi]$ is an initial data vector with respect to $\xi$ only, $U[\xi,\phi]$ is a unitary evolution operator
\begin{eqnarray}
U=\exp\left\{-i\int_{\Sigma(\xi)}\delta\xi' H[\xi',\phi]\right\}=\exp\left\{-i\Omega(\xi,\phi) \langle H\rangle(\xi,\phi)\right\},
\end{eqnarray}
and $\Sigma(\xi)$ is a finite integration area in $\xi$-space, whereas the volume $\Omega$ of full configuration space and the averaged energy $\langle H\rangle$ are
\begin{equation}\label{formulas}
\Omega(\xi,\phi)=\int_{\Sigma(\xi,\phi)}\delta\xi'\delta\phi'\quad,\quad\langle H\rangle(\xi,\phi)=\dfrac{1}{\Omega(\xi,\phi)}\int_{\Sigma(\xi)}\delta\xi' H[\xi',\phi].
\end{equation}
where $\Sigma(\xi,\phi)=\Sigma(\xi)\times\Sigma(\phi)$ is a finite integration region of full configuration space. Explicitly
\begin{eqnarray}
  &&U[\xi,\phi]=\mathbf{1}_2\cosh\left[\Omega(\xi,\phi)\sqrt{\strut{\langle V\rangle(\xi,\phi)}}\right]+\\
  &&+\left[\begin{array}{cc}0&\sqrt{\strut{\langle V\rangle(\xi,\phi)}}\\ \left(\sqrt{\strut{\langle V\rangle(\xi,\phi)}}\right)^{-1}&0\end{array}\right]\sinh\left[\Omega(\xi,\phi)\sqrt{\strut{\langle V\rangle(\xi,\phi)}}\right],
\end{eqnarray}
where
\begin{equation}
  \langle V\rangle(\xi,\phi)=\dfrac{1}{\Omega(\xi,\phi)}\int_{\Sigma(\xi)}\delta\xi'V[\xi',\phi],
\end{equation}
and, consequently, the received wave functional are
\begin{eqnarray}\label{wfun}
  \Psi[\xi,\phi]&=&\Psi[\xi^I,\phi]\cosh\left[\Omega(\xi,\phi)\sqrt{\strut{\langle V\rangle(\xi,\phi)}}\right]+\nonumber\\
  &+&\Pi_\Psi[\xi^I,\phi]\left(\sqrt{\strut{\langle V\rangle(\xi,\phi)}}\right)^{-1}\sinh\left[\Omega(\xi,\phi)\sqrt{\strut{\langle V\rangle(\xi,\phi)}}\right],
\end{eqnarray}
whereas the canonical conjugate momentum is
\begin{eqnarray}\label{pfun}
  \Pi_\Psi[\xi,\phi]&=&\Pi_\Psi[\xi^I,\phi]\cosh\left[\Omega(\xi,\phi)\sqrt{\strut{\langle V\rangle(\xi,\phi)}}\right]+\nonumber\\
  &+&\Psi[\xi^I,\phi]\sqrt{\strut{\langle V\rangle(\xi,\phi)}}\sinh\left[\Omega(\xi,\phi)\sqrt{\strut{\langle V\rangle(\xi,\phi)}}\right],
\end{eqnarray}
where $\Psi[\xi^I,\phi]$ and $\Pi_\Psi[\xi^I,\phi]$ are initial data with respect to $\xi$. Applying (\ref{1}) in (\ref{pfun}), one obtains
\begin{eqnarray}\label{pi1}
\!\!\!\!\!\!\!\!\!\!&&\Pi_\Psi[\xi,\phi]=\dfrac{\Pi_\Psi[\xi^I,\phi]}{\sqrt{\strut{\langle V\rangle}}}\dfrac{\delta}{\delta\xi}\left[\Omega\sqrt{\strut{\langle V\rangle}}\right]\cosh\left[\Omega\sqrt{\strut{\langle V\rangle}}\right]+\nonumber\\
\!\!\!\!\!\!\!\!\!\!&&+\left[\Psi[\xi^I,\phi]\dfrac{\delta}{\delta\xi}\left[\Omega\sqrt{\strut{\langle V\rangle}}\right]+\Pi_\Psi[\xi^I,\phi]\dfrac{\delta}{\delta\xi}\left[\left(\sqrt{\strut{\langle V\rangle}}\right)^{-1}\right]\right]\sinh\left[\Omega\sqrt{\strut{\langle V\rangle}}\right],~~~~~~~
\end{eqnarray}
where $\Omega\equiv\Omega(\xi,\phi)$ and $\langle V\rangle\equiv\langle V\rangle(\xi,\phi)$, and calculating
\begin{eqnarray}
\!\!\!\!\!\!\!\!\!\!\dfrac{\delta}{\delta\xi}\left[\Omega\sqrt{\strut{\langle V\rangle}}\right]&=&\dfrac{1}{2}\sqrt{\strut{\langle V\rangle}}\left(\dfrac{\delta\Omega}{\delta\xi}+1\right),\\
\!\!\!\!\!\!\!\!\!\!\dfrac{\delta}{\delta\xi}\left[\left(\sqrt{\strut{\langle V\rangle}}\right)^{-1}\right]&=&\dfrac{1}{2}\left[\Omega\sqrt{\strut{\langle V\rangle}}\right]^{-1}\left(\dfrac{\delta\Omega}{\delta\xi}-1\right),
\end{eqnarray}
with using (\ref{pi1}), one receives the formula
\begin{eqnarray}\label{pi2}
\!\!\!\!\!\!\!\!\!\!&&\Pi_\Psi[\xi,\phi]=\Pi_\Psi[\xi^I,\phi]\dfrac{1}{2}\left(\dfrac{\delta\Omega}{\delta\xi}+1\right)\cosh\left[\Omega\sqrt{\strut{\langle V\rangle}}\right]+\nonumber\\
\!\!\!\!\!\!\!\!\!\!&&+\left[\Psi[\xi^I,\phi]\dfrac{\sqrt{\strut{\langle V\rangle}}}{2}\left(\dfrac{\delta\Omega}{\delta\xi}+1\right)+\dfrac{\Pi_\Psi[\xi^I,\phi]}{2\Omega\sqrt{\strut{\langle V\rangle}}}\left(\dfrac{\delta\Omega}{\delta\xi}-1\right)\right]\sinh\left[\Omega\sqrt{\strut{\langle V\rangle}}\right],~~~~~~~
\end{eqnarray}
which compared with (\ref{pfun}) leads to the system of equations
\begin{eqnarray}\label{seq}
\left\{\begin{array}{cc}\dfrac{1}{2}\left(\dfrac{\delta\Omega}{\delta\xi}+1\right)=1\vspace*{10pt}\\
\Psi[\xi^I,\phi]\dfrac{1}{2}\left(\dfrac{\delta\Omega}{\delta\xi}+1\right)+\dfrac{\Pi_\Psi[\xi^I,\phi]}{\Omega\langle V\rangle}\dfrac{1}{2}\left(\dfrac{\delta\Omega}{\delta\xi}-1\right)=\Psi[\xi^I,\phi]\end{array}\right..
\end{eqnarray}
The first equation of the system (\ref{seq}) yields the relation
\begin{equation}
\dfrac{\delta\Omega}{\delta\xi}=1=\int_{\Sigma(\phi)}\delta\phi',
\end{equation}
where the last integral arises by the first formula in (\ref{formulas}), which after application to the second equation gives simply $\Psi[\xi^I,\phi]=\Psi[\xi^I,\phi]$ and, moreover, the volume $\Omega(\xi,\phi)$ is $\phi$-invariant
\begin{equation}
  \Omega(\xi,\phi)=\int_{\Sigma(\xi)}\delta\xi'=\Omega(\xi).
\end{equation}

The probability density can be deduced easily by (\ref{wfun})
\begin{eqnarray}
  |\Psi[\xi,\phi]|^2&=&(\Psi[\xi^I,\phi])^2\cosh^2\left[\Omega\sqrt{\langle V\rangle}\right]+\nonumber\\
  &+&(\Pi_\Psi[\xi^I,\phi])^2\left(\langle V\rangle\right)^{-1}\sinh^2\left[\Omega\sqrt{\langle V\rangle}\right]+\nonumber\\
  &+&\Psi[\xi^I,\phi]\Pi_\Psi[\xi^I,\phi]\left(\sqrt{\langle V\rangle}\right)^{-1}\sinh\left[2\Omega\sqrt{\langle V\rangle}\right],
\end{eqnarray}
and, in the light of the relation (\ref{coord}), one has
\begin{equation}
  |\Psi[\xi,\phi]|^2=(\Psi[\xi^I,\phi])^2\cosh^2\left[\Omega\sqrt{\langle V\rangle}\right]+(\Pi_\Psi[\xi^I,\phi])^2\left(\langle V\rangle\right)^{-1}\sinh^2\left[\Omega\sqrt{\langle V\rangle}\right].
\end{equation}
Assuming the following separation conditions
\begin{eqnarray}\label{sep}
  \Psi[\xi^I,\phi]=\Psi[\xi^I]\Gamma_\Psi[\phi]\quad,\quad\Pi_\Psi[\xi^I,\phi]=\Pi_\Psi[\xi^I]\Gamma_\Pi[\phi],
\end{eqnarray}
where $\Gamma_\Psi$ and $\Gamma_\Pi$ are functionals of $\phi$ only, while $\Psi[\xi^I]$ and $\Pi_\Psi[\xi^I]$ are constant functionals, and applying the usual normalization, one obtains
\begin{equation}\label{abc}
  \int_{\Sigma(\xi,\phi)}|\Psi[\xi',\phi']|^2\delta\xi'\delta\phi'=1\longrightarrow A(\Pi_\Psi[\xi^I])^2+B(\Psi[\xi^I])^2-1=0,
\end{equation}
where the constants $A$ and $B$ are given by the integrals
\begin{eqnarray}
  A&=&\int_{\Sigma(\xi,\phi)}\Gamma_\Pi[\phi']\left(\langle V'\rangle\right)^{-1}\sinh^2\left[\Omega'\sqrt{\langle V'\rangle}\right]\delta\xi'\delta\phi',\\
  B&=&\int_{\Sigma(\xi,\phi)}\Gamma_\Psi[\phi']\cosh^2\left[\Omega'\sqrt{\langle V'\rangle}\right]\delta\xi'\delta\phi',
\end{eqnarray}
assumed to be convergent and finite. The solution to the equation (\ref{abc})
\begin{eqnarray}\label{pii}
\Pi_\Psi[\xi^I]=\pm\sqrt{\strut{\dfrac{1}{A}-\dfrac{B}{A}(\Psi[\xi^I])^2}},
\end{eqnarray}
joined with (\ref{1}) and (\ref{sep}) gives the differential equation for the initial data
\begin{equation}\label{de}
  \dfrac{1}{\Gamma[\phi]}\dfrac{\delta \Psi[\xi^I]}{\delta \xi^I}=\pm\sqrt{\strut{\dfrac{1}{A}-\dfrac{B}{A}(\Psi[\xi^I])^2}},\quad \Gamma[\phi]\equiv\dfrac{\Gamma_\Pi[\phi]}{\Gamma_\Psi[\phi]},
\end{equation}
which can be integrated
\begin{equation}
  \sqrt{A}\int\dfrac{\delta \Psi[\xi^I]}{\sqrt{\strut{1-B(\Psi[\xi^I])^2}}}=\pm\Gamma[\phi]\xi^I+C,
\end{equation}
where $C$ is a constant of integration, and gives the formula
\begin{equation}
  \sqrt{\strut{A/B}}\arcsin\left\{\sqrt{\strut{B/A}}\Psi[\xi^I]\right\}=\pm\Gamma[\phi]\xi^I+C,
\end{equation}
which is equivalent to
 \begin{equation}
 \Psi[\xi^I]=\sqrt{\strut{A/B}}\sin\theta(\xi^I,\phi),\quad \theta(\xi^I,\phi)=\sqrt{\strut{B/A}}\left(\pm\Gamma[\phi]\xi^I+C\right.
 \end{equation}
Because $\Psi[\xi^I]$ must be a functional of $\xi^I$, one has $\Gamma[\phi]=\Gamma_0$ with a constant $\Gamma_0$, and, moreover, $\theta(\xi^I,\phi)=\theta(\xi^I)$. Taking into account (\ref{pii}), one obtains
 \begin{eqnarray}
\Psi[\xi^I]=\sqrt{\strut{A/B}}\sin\theta(\xi^I)\quad,\quad\Pi_\Psi[\xi^I]=\pm\sqrt{\strut{\dfrac{1}{A}-\sin^2\theta(\xi^I)}}.
 \end{eqnarray}
 In the light of the equation (\ref{coord}), however, one of the relations is always true
 \begin{eqnarray}
\sin\theta(\xi^I)\equiv0\quad,\quad\sin\theta(\xi^I)=\pm\sqrt{\strut{1/A}}.\label{1st}
 \end{eqnarray}
One sees that in any case $\xi_I$ has discrete values. By the first relation in (\ref{1st})
 \begin{equation}
   \sqrt{\strut{B/A}}\left(\pm\Gamma_0\xi^I+C\right)=k\pi \longrightarrow \xi^I=\pm\dfrac{1}{\Gamma_0}\left(\sqrt{\strut{A/B}}k\pi-C\right),
 \end{equation}
 where $k\in \mathbb{Z}$ is an integer, while the second relation in (\ref{1st}) gives
 \begin{equation}
 \xi^I=\pm\dfrac{1}{\Gamma_0}\left(\pm\sqrt{\strut{A/B}}\arcsin\sqrt{\strut{1/A}}-C\right).
 \end{equation}
 For the first case one has
 \begin{equation}
   \Psi[\xi^I]=0\quad,\quad \Pi_\Psi[\xi^I]=\pm\sqrt{\strut{1/A}},
 \end{equation}
 whereas and for the second one
 \begin{eqnarray}
   \Psi[\xi^I]=\pm\sqrt{\strut{1/B}}\quad,\quad \Pi_\Psi[\xi^I]=0.
 \end{eqnarray}
 Finally, the invariant one-dimensional wave functional (\ref{wfun}) becomes
 \begin{equation}\label{sol1}
 \Psi[\xi,\phi]=\pm\Gamma_\Psi[\phi]\Gamma_0\sqrt{\strut{\dfrac{1}{A}}}\left(\sqrt{\strut{\langle V\rangle(\xi,\phi)}}\right)^{-1}\sinh\left[\Omega(\xi)\sqrt{\strut{\langle V\rangle(\xi,\phi)}}\right],
 \end{equation}
 in the first case of (\ref{1st}), while for the second one
 \begin{equation}\label{sol2}
\Psi[\xi,\phi]=\pm\Gamma_\Psi[\phi]\sqrt{\strut{\dfrac{1}{B}}}\cosh\left[\Omega(\xi)\sqrt{\strut{\langle V\rangle(\xi,\phi)}}\right].
\end{equation}
\section{Developments}\label{sec:5}
\subsection{General solutions}
The general analytic solutions of the reduced quantum geometrodynamics can be now constructed for any induced metric $h_{ij}$ from the solutions (\ref{sol1}) and (\ref{sol2}). It can be easily seen that
\begin{equation}\label{formula}
\langle V\rangle(h_{ij},\phi)=\langle{^{(3)}\!R[h]}\rangle-2\Lambda-6\langle\rho[\phi]\rangle,
\end{equation}
where
\begin{eqnarray}
\langle{^{(3)}\!R[h]}\rangle&=&\dfrac{1}{\Omega(h_{ij})}\int_{\Sigma(h_{ij})}\delta h_{ij}'\sqrt{\dfrac{2}{3}}\sqrt{\strut{h'}}{h^{ij}}'~{^{(3)}\!R[h']},\\
\langle\rho[\phi]\rangle&=&\int_{\Sigma(\phi)} \delta \phi'\rho[\phi'],
\end{eqnarray}
and
\begin{equation}
  \Omega(h_{ij})=\int_{\Sigma(h_{ij})}\delta h_{ij}'\sqrt{\dfrac{2}{3}}\sqrt{\strut{h'}}{h^{ij}}'.
\end{equation}
Making use of (\ref{formula}) in the solutions (\ref{sol1}) and (\ref{sol2}), one obtains the general solutions according to the global one-dimensionality conjecture
\begin{eqnarray}
\!\!\!\!\!\!\!\!\!\!&&\Psi[h_{ij},\phi]=\pm\Gamma_\Psi[\phi]\Gamma_0\sqrt{\strut{\dfrac{1}{A}}}\left(\sqrt{\strut{\langle V\rangle(h_{ij},\phi)}}\right)^{-1}\sinh\left[\Omega(h_{ij})\sqrt{\strut{\langle V\rangle(h_{ij},\phi)}}\right],~~~~\label{solm1}\\
\!\!\!\!\!\!\!\!\!\!&&\Psi[h_{ij},\phi]=\pm\Gamma_\Psi[\phi]\sqrt{\strut{\dfrac{1}{B}}}\cosh\left[\Omega(h_{ij})\sqrt{\strut{\langle V\rangle(h_{ij},\phi)}}\right],~~~~\label{solm2}
\end{eqnarray}
where
\begin{eqnarray}
\!\!\!\!\!\!\!\!\!\!&&A=\sqrt{\strut{\dfrac{2}{3}}}\Gamma_0\int_{\Sigma(h_{ij},\phi)}\Gamma_\Psi[\phi']\dfrac{\sinh^2\left[\Omega(h_{ij}')\sqrt{\langle V\rangle(h_{ij}',\phi')}\right]}{\langle V\rangle(h_{ij}',\phi')}\sqrt{\strut{h'}}{h^{ij}}'\delta h_{ij}'\delta\phi',~~~~~~~~~~\\
\!\!\!\!\!\!\!\!\!\!&&B=\sqrt{\strut{\dfrac{2}{3}}}\int_{\Sigma(h_{ij},\phi)}\Gamma_\Psi[\phi']\cosh^2\left[\Omega(h_{ij}')\sqrt{\langle V\rangle(h_{ij}',\phi')}\right]\sqrt{\strut{h'}}{h^{ij}}'\delta h_{ij}'\delta\phi',~~~~~~~~~~
\end{eqnarray}
are assumed to be convergent and finite constants. The normalization condition
\begin{equation}\label{normaliz}
  \int_{\Sigma(h_{ij},\phi)}|\Psi[h_{ij},\phi]|^2\sqrt{\dfrac{2}{3}}\sqrt{\strut{h'}}{h^{ij}}'\delta h_{ij}'\delta\phi=1,
\end{equation}
applied in the solutions (\ref{solm1}) and (\ref{solm2}), leads to
\begin{equation}
  |\Gamma_\Psi[\phi]\Gamma_0|^2=1\quad,\quad \Gamma_\Psi[\phi]\Gamma_0=1,
\end{equation}
which yield $\Gamma_\Psi[\phi]=1/\Gamma_0$, $\Gamma_0=1$ and, therefore,
\begin{eqnarray}
\!\!\!\!\!\!\!\!\!\!&&\Psi_1[h_{ij},\phi]=\pm\sqrt{\strut{\dfrac{1}{A}}}\left(\sqrt{\strut{\langle V\rangle(h_{ij},\phi)}}\right)^{-1}\sinh\left[\Omega(h_{ij})\sqrt{\strut{\langle V\rangle(h_{ij},\phi)}}\right],~~~~~~~~\label{solmi1}\\
\!\!\!\!\!\!\!\!\!\!&&\Psi_2[h_{ij},\phi]=\pm\sqrt{\strut{\dfrac{1}{B}}}\cosh\left[\Omega(h_{ij})\sqrt{\strut{\langle V\rangle(h_{ij},\phi)}}\right],~~~~~~~~\label{solmi2}
\end{eqnarray}
where
\begin{eqnarray}
\!\!\!\!\!\!\!\!\!\!&&A=\sqrt{\strut{\dfrac{2}{3}}}\int_{\Sigma(h_{ij},\phi)}\dfrac{\sinh^2\left[\Omega(h_{ij}')\sqrt{\langle V\rangle(h_{ij}',\phi')}\right]}{\langle V\rangle(h_{ij}',\phi')}\sqrt{\strut{h'}}{h^{ij}}'\delta h_{ij}'\delta\phi',~~~~~~~~~~\\
\!\!\!\!\!\!\!\!\!\!&&B=\sqrt{\strut{\dfrac{2}{3}}}\int_{\Sigma(h_{ij},\phi)}\cosh^2\left[\Omega(h_{ij}')\sqrt{\langle V\rangle(h_{ij}',\phi')}\right]\sqrt{\strut{h'}}{h^{ij}}'\delta h_{ij}'\delta\phi'.~~~~~~~~~~
\end{eqnarray}
The solutions (\ref{solmi1}) and (\ref{solmi2}) describe two independent quantum gravity states.
\subsection{Superposition}
Because, the equations (\ref{wdw}) and (\ref{eqn}) are linear, the superposition
\begin{eqnarray}
\!\!\!\!\!\!\!\!\!\!&&\Psi=\sum_{i=1,2}\alpha_i\Psi_i\label{solmi2X}
\end{eqnarray}
where $\alpha_i$ are arbitrary constants, could be considered as the most general solution, for which the normalization condition (\ref{normaliz}) is the constraint
\begin{equation}\label{relc}
|\alpha_1|^2+|\alpha_2|^2+(\alpha^\star_1\alpha_2+\alpha_1\alpha^\star_2)I=1,
\end{equation}
where
\begin{equation}
  I=\sqrt{\strut{\dfrac{1}{AB}}}\int_{\Sigma(h_{ij},\phi)} \dfrac{\sinh\left[2\Omega(h_{ij}')\sqrt{\strut{\langle V\rangle(h_{ij}',\phi')}}\right]}{2\sqrt{\strut{\langle V\rangle(h_{ij}',\phi')}}}\sqrt{\dfrac{2}{3}}\sqrt{\strut{h'}}{h^{ij}}'\delta h_{ij}'\delta\phi'.
\end{equation}
For $I=0$, (\ref{relc}) gives simply
\begin{equation}
  |\alpha_2|=\sqrt{\strut{1-|\alpha_1|^2}}\quad,\quad|\alpha_1|\geqslant1.
\end{equation}
The case $I\neq0$ is more complicated. Note that the constraint (\ref{relc}) gives
\begin{equation}
  (\alpha_1+\alpha_2I)\alpha^\star_1+(\alpha_2+\alpha_1I)\alpha_2^\star=0\longrightarrow\dfrac{\alpha^\star_1}{\alpha_2^\star}=\dfrac{-\alpha_1I+\alpha_2}{\alpha_1+\alpha_2I},
\end{equation}
or, equivalently, for $0\neq C\in\mathbb{R}$
\begin{equation}\label{CC}
C\alpha^\star_1=-\alpha_1I+\alpha_2\quad,\quad C\alpha_2^\star=\alpha_1+\alpha_2 I,
\end{equation}
and
\begin{equation}
C|\alpha_1|^2=-\alpha^2_1 I+\alpha_2\alpha_1\quad,\quad C|\alpha_2|^2=\alpha_1\alpha_2+\alpha_2^2 I,
\end{equation}
which after mutual adding and making use of (\ref{relc}) gives
\begin{equation}
  CI[(\alpha^\star_1-\alpha_2)\alpha_2+(\alpha_2^\star+\alpha_1)\alpha_1]=\alpha_1\alpha_2+\alpha_2\alpha_1,
\end{equation}
and, consequently,
\begin{equation}\label{in}
CI(\alpha^\star_1-\alpha_2)=\alpha_1\quad,\quad CI(\alpha_2^\star+\alpha_1)=\alpha_2.
\end{equation}
The complex decomposition for $\alpha$ and $\alpha_2$ applied in (\ref{in}) leads to
\begin{equation}
\Re\alpha_2=(CI-1)\Re\alpha_1\quad,\quad\Im\alpha_2=(CI-1)\Im\alpha_1,
\end{equation}
or, equivalently,
\begin{equation}\label{eqi}
\alpha_2=(CI-1)\alpha_1\quad,\quad|\alpha_2|^2=(CI-1)^2|\alpha_1|^2.
\end{equation}
Employing (\ref{eqi}) within the constraint (\ref{relc}), one obtains
\begin{equation}\label{alp}
|\alpha_1|^{-2}=IC^2+(I^2-2I)C-I+2.
\end{equation}
Because both $|\alpha_i|^2\in\mathbb{R}$ as squares of absolute values, one obtains the values of the constant $C$ in dependence on the integral $I$
\begin{equation}
C\in[-\infty,C_-]\cup[C_+,\infty]\quad,\quad C_{\pm}=\dfrac{I-2}{2}\left[1\pm\sqrt{\strut{1-\dfrac{4}{I(I-2)}}}\right],
\end{equation}
where for $C_\pm\in\mathbb{R}$ the condition $I\in[-\infty,1-\sqrt{\strut{5}}]\cup[1+\sqrt{\strut{5}},\infty]$ holds.

\section{Summary}\label{sec:6}
We have discussed few consequences of quantum geometrodynamics according to the global one-dimensional conjecture. Employment of the conjecture immediately led us to construction of the analytic solutions, wherein the strategy of integration used the concept of invariant dimension instead of the global dimension introduced to remove the singular behavior of the effective potential. In general, the procedure has used for computations the Lebesgue--Stieltjes, or Radon, one-dimensional integrals, and, therefore, meaningfully simplified considerations of quantum gravity and led to analytical wave functionals. Finally, we have discussed developments of the strategy. The first one was construction of the solutions for any induced metric, which differ from the Feynman path integral solutions, whereas the second one was the question of superposition. Certainly, there are open problems related to the novel wave functionals. The reader interested in advancements is advised to take into account the author's monograph \cite{glinkamon}.

\end{document}